\documentclass[12pt,preprint]{aastex}
\def\gtsima{$\, \buildrel > \over \sim \,$}
\def\ltsima{$\, \buildrel < \over \sim \,$}
\def\simgt{\lower.5ex\hbox{\gtsima}}
\def\simlt{\lower.5ex\hbox{\ltsima}}
 
\begin{document}
 
\title{$K$-Band Red Clump Distance to the Large Magellanic Cloud}

\author{David R. Alves}
\affil{Columbia Astrophysics Laboratory, 
550 W. 120th St., MailCode 5247, NY, NY, 10027 \\ 
Email: {\tt alves@astro.columbia.edu} }

\author{Marina Rejkuba\altaffilmark{1}}
\affil{European Southern Observatory, 
Karl-Schwarzschild-Strasse 2, 85748 Garching, Germany \\
Email: {\tt mrejkuba@eso.org} }

\altaffiltext{1}{Depto.~de Astronomia, P.~Universidad Catolica,
Casilla 306, Santiago 22, Chile }

\author{Dante Minniti}
\affil{Depto.~de Astronomia, P.~Universidad Catolica,
Casilla 306, Santiago 22, Chile \\
Email: {\tt dante@astro.puc.cl} }

\author{Kem H. Cook}
\affil{Lawrence Livermore National Laboratory, 
Livermore, CA 94550 \\
Email: {\tt kcook@llnl.gov } }

\begin{abstract}
The {\it Hipparcos} $I$-band calibration 
of horizontal-branch red clump giants 
as standard candles has lead to controversial
results for the distance to the Large Magellanic Cloud (LMC).
In an attempt to properly ascertain
the corrections for interstellar extinction and
clump age and metallicity,
we analyze new multi-wavelength luminosity functions
of the LMC red clump.
Our photometry dataset in the $K$-band 
was obtained with the SOFI infrared imager
at the European Southern Observatory's New Technology Telescope.
In the $V$ and $I$ passbands, we employ data from
WFPC2 onboard the {\it Hubble Space Telescope\,}.
The LMC red clump is first identified in a
$K$,$(V-K)$ color-magnitude diagram.
Our luminosity functions 
yield apparent magnitudes 
of $K$ = 16.974, $I$ = 18.206, and
$V$ = 19.233 (\,$\pm 0.009_r \pm 0.02_s$; random and
systematic error, respectively).
Compared directly to the {\it Hipparcos\,} red clump calibration
(without a correction for age and metallicity),
the LMC clump measurements
imply a negative interstellar reddening correction.
This unphysical result indicates a population difference between clumps.
A modified calibration
based on theoretical modeling yields
an average reddening correction of
$E(B-V) = 0.089 \pm 0.015_r$, and
a true LMC distance modulus of $\mu_0 = 18.493 \pm 0.033_r
\pm 0.03_s$.  We reconcile our result with the 
short distance previously derived from OGLE~II red clump data.

\end{abstract}


\section{Introduction}

The {\it Hipparcos\,} satellite mission
yielded parallax measurements
for hundreds of nearby red clump stars, and these are, collectively,
the best calibrated standard candle
(Paczy\'{n}ski \& Stanek 1998; Paczy\'{n}ski~2001).
However, attempts to use
the {\it Hipparcos\,} $I$-band red clump calibration
to determine the
distance to the Large Magellanic Cloud (LMC)
have lead to controversial results.
Udalski et al.~(1998) obtained $\mu_0 = 18.08 \pm 0.03_r \pm 0.12_s$ 
(random and systematic error, respectively),
which was subsequently revised upward
to $\mu_0 = 18.24 \pm 0.08$ (Udalski 2000).
Stanek, Zaritsky and Harris~(1998) found
$\mu_0 = 18.07 \pm 0.03_r \pm 0.09_s$.
However, 
Zaritsky (1999) showed that the reddening correction
used by Stanek et al.~(1998) was overestimated.
Romaniello et al.~(2000) found
$\mu_0 = 18.59 \pm 0.04_r \pm 0.08_s$
by analyzing {\it Hubble Space Telescope\,} ({\it HST\,}) photometry
of the LMC red clump.
Girardi and Salaris (2001) review
these measurements, and apply new
model corrections for population
effects (i.e.~age and metallicity).
They find a population correction 
in the $I$-band of 0.2~mag.  
Consequently, they revise the Udalski (2000) 
result to $\mu_0 = 18.37 \pm 0.07$, and the
Zaritsky (1999) and Romaniello et al.~(2000) result
to $\mu_0 = 18.55 \pm 0.05$.

Alves~(2000) demonstrated the use of the red clump as a standard candle 
at longer wavelengths; he calibrated the
{\it Hipparcos} red clump absolute magnitude in the $K$-band,
and applied the calibration to the Galactic bulge red clump.
One advantage of working in
this $\sim$2$\micron$ passband is the particularly small effect
of extinction by interstellar dust 
(e.g.~Schlegel et al.~1998).
Grocholski and Sarajedini (2002) recently studied 
population effects on the red clump's $K$-band absolute magnitude
by compiling 2MASS photometry of open clusters.
They concluded that $M_K$ is, in general, also less sensitive
to population effects than $M_I$.
The red clump models 
of Girardi and Salaris (2001)
were also shown to be in fair agreement 
with the $K$-band open cluster data.
In this {\it Letter\,},
we present a deep $K$-band 
luminosity function for red clump stars
in the LMC bar and inner-disk.
We also construct a $K$, $(V-K)$ color-magnitude
diagram, and complementary $V$ and $I$ luminosity functions.
This is a first reconnaissance of optical
and near-infrared photometry of the LMC red clump.

\section{Data}

The 6 lines-of-sight studied in this paper are centered on
the source stars of LMC microlensing events 
discovered by the MACHO Project (Alcock et al.~2001;
events 4, 7, 9, 12, 14, and 18).  
For our purposes,
the lines-of-sight
were selected at random by the natural occurrence of
microlensing\footnote{Actually, event 12 may be a 
supernova in a galaxy behind the LMC, but this
distinction does not affect the current work.}.
The 6-field average line-of-sight intersects the LMC disk
at $\alpha$ = 05:25:55.2 and $\delta$ = $-$70:20:24~(J2000).
This is $0^{\circ}.49$ from the center of the LMC, 
nearly perpendicular to the line-of-nodes, and
on the far side of the inclined disk (van~der~Marel et al.~2002).
The geometric correction 
is small, about 0.6\% in distance (0.013 mag), and in
the sense that our clump giants are farther away
than those at the LMC center.

The $K$-band dataset analyzed here was obtained with
the SOFI infrared imager
at the European Southern Observatory's New Technology Telescope;
it is a 1024$\times$1024 array with a pixel
size of 0$^{\prime\prime}$.292.  The total field-of-view
of our trimmed (equal-exposure) images after coadding
individual dithered frames is about 140 $\sq^{\,\prime}$.
The photometry was derived with DAOPHOT II (Stetson 1994),
and calibrated with observations of 6 standard stars (Persson et al.~1998).
Conditions were photometric (Rejkuba et al.~2001).
The 1-$\sigma$ standard deviation of the calibration solution
is 0.028~mag.  
Over 98\% of the stars in the red clump 
(or brighter; $K < 18$)
were detected with a signal-to-noise $>$ 10, and
thus our SOFI red clump data are nearly complete.
The $K_s$ mag system employed (``K-short'')
is the same as 2MASS.
In order to check our standardized zero point, we 
matched a subset of the SOFI data
with 2MASS second incremental release 
photometry\footnote{{\tt http://www.ipac.caltech.edu/2mass/}.}.
A comparison 
shows no zero-point difference to within 0.02~mag
(for stars with colors like the red clump).
However, 
all of the stars in the comparison are actually redder and
brighter than the clump, and thus the 
agreement found lends only weak support to the
accuracy of our data.
Finally, $K_s$~magnitudes are related to
the $K$-band red clump data compiled by Alves (2000) 
by $K = K_s + 0.044$ (Grocholski \& Sarajedini 2002;
their Eqn.~2).
We transform $K_s$ to $K$ accordingly.

Our dataset in $V$ and $I$ was obtained
with WFPC2 onboard the {\it HST\,} 
(PI:~Cook).
We exclude the PC2 images from our analysis.
This guarantees a homogeneous $V$ and $I$ photometry dataset,
and represents only a small loss in imaged area.
The 3 WFs are 800$\times$800 pixel arrays
with a pixel size of 0$^{\prime\prime}$.1.
The resulting total field-of-view 
is approximately 32~$\sq^{\,\prime}$.
The {\it HST\,}/WF instrumental F555W and F814W 
magnitudes were derived with DAOPHOT~II, and calibrated to 
Landolt's $V$ and $I$ with an accuracy of 0.02--0.03~mag 
following the usual procedures (Alcock et al.~2001).
Nearly all of the LMC clump giants were detected
with very high signal-to-noise.
In order to check our $V$ and $I$ zero points, 
we cross-correlated a subset of our data
with OGLE~II ground-based 
photometry\footnote{{\tt http://sirius.astrouw.edu.pl/ogle/}.}
(Udalski et al.~2000).
A comparison reveals zero-point differences of $\Delta V = 0.11$ 
and $\Delta I = 0.02$~mag 
in the sense that OGLE~II is brighter.  
Alcock et al.~(1999)
compared MACHO ground-based photometry with an
early reduction of these same 
standardized {\it HST\,}/WF data (a different subset),
and found that MACHO was brighter 
by $\Delta V = 0.06$ and
$\Delta R = 0.00$ mag.
We speculate that the aperture corrections applied
to the OGLE~II and MACHO ground-based $V$ data 
may be affected at the 5-10\% level
by crowding errors.  For example, it is possible
that the $V$-band ``sky''
around bright stars used to determine
the aperture corrections was on average underestimated
because nearby neighbor stars were oversubtracted
(Alcock et al.~1999; Udalski et al.~2000).
We tentatively adopt a formal systematic 
calibration error of
0.02~mag in $V$ and $I$ based on the
stated accuracy of {\it HST\,}/WF calibrations.

\section{Analysis and Results }

We detected approximately 36,000 stars in $K$ and
109,000 stars in $V$ and $I$.  Of these approximately
27,000 have $K < 20$, and 7,000 have
$V < 21$.  Cross-correlating these latter source lists
yields 4745 stars with $K$, $I$, and $V$ mags.
Figure 1 shows the resulting
$K$,$(V-K)$ color-magnitude diagram (CMD)
in the region around the red clump.  Only 2353 stars appear
within the limits of Fig.~1, and we restrict our
subsequent analyses to these stars.
As discussed by Alves (2000; and refs.~therein), not all
of the stars appearing in Fig.~1 are bona~fide red clump
giants.  The first-ascent red giant branch (RGB) is
identified as the roughly vertical sequence running from
$(V-K) \sim$ 2.2 at $K \sim 18.5$ to $(V-K) \sim$ 3.1
at $K \sim 14.5$.  It has a width of about 0.2 mag in
$(V-K)$ color, and no distinct component branches are evident.
Some of these stars are second-ascent
asymptotic giant branch (AGB) stars.   The distinction between
an RGB and AGB star is not necessarily clear in a CMD of
mixed-age field populations like this one.
The sought-after horizontal branch red clump 
is the most prominent
feature in Fig.~1.  It appears at $K \sim 17$ and lies
mostly blueward of the giant branch, i.e.~$(V-K) \simlt 2.4$.
The detailed structure of the clump seen here
is well understood in the context of
stellar evolution theory (Girardi \& Salaris 2001).
The overdensity of stars with $K \sim 17$ and colors
that associate them with the giant branch 
(i.e.~$V-K \sim 2.4$ to 2.5) may be the first detection
of the ``RGB bump'' for LMC field stars
(see Fig.~19 of Girardi \& Salaris 2001).
The RGB bump
is an evolutionary pause as the hydrogen-burning shell
crosses a chemical composition discontinuity.
The loose clustering of stars seen on the giant branch at 
$(V-K) \sim 2.8$ and $K \sim 15.5$ may be 
the ``AGB bump,'' an evolutionary pause at the onset of
helium shell burning
(e.g., see Alves \& Sarajedini 1999).

A side-by-side comparison of 
red clump luminosity functions (LFs)
is presented in Figure~2.
The LMC clump is shown on the left.  The {\it Hipparcos\,} 
clump is shown on the right (data from Alves~2000).
The LFs in $K$, $I$, and $V$ are each plotted 
in separate panels (top to bottom).
LMC bin size is 0.05 mag; {\it Hipparcos\,}
bin size is 0.10 mag; each panel is 1.5~mag wide.
The adopted model LFs (solid lines in Fig.~2)
are a superposition
of a linear background and a Gaussian of variable width,
amplitude, and location.
The backgrounds are fit in 
restricted magnitude ranges which exclude the clump.
The best-fit model parameters
and $\chi^2$ (per d.o.f.) are provided in Table~1.
The LMC red clump in Fig.~2 totals about 2000 stars.
For comparison, there are 238 {\it Hipparcos\,} red clump
stars with $K$, $I$, and $V$ data (Alves 2000).
This distinction reflects in the formal statistical
errors for the peak clump brightnesses given in Table~1.
Despite the shortcomings 
of our model LFs revealed in Fig.~2,
the peak brightnesses of the red clumps
are evidently well determined.

The {\it Hipparcos\,} calibration assumes no reddening
correction, as suggested by Stanek and Garnavich (1998).
These authors argue that the average reddening is
probably $E(B-V) < 0.02$~mag.  It should
be kept in mind that the LMC distance derived in
this work is practically
independent of the reddening correction
applied to the {\it Hipparcos} clump.
However, the average LMC reddening obtained 
in comparison may be slightly underestimated.
We note that the Lutz-Kelker correction to the Alves (2000)
calibration dataset is probably also negligible (Udalski 2000).
The true LMC distance modulus is therefore given by:
\begin{equation}
\mu_0 \ = \ (m-M)_{0} \ = \ (m -  M)_{\lambda} \ - \ 
A_{\lambda} \ + \ \Delta M_{\lambda} 
\end{equation}
where $\lambda$ corresponds to either $K$, $I$, or $V$,
and the apparent distance moduli in each passband are
$(m-M)_{\lambda}$.
The corrections for population
effects are $\Delta M_{\lambda} =
M_{\lambda}(Hipp.) - M_{\lambda}$(LMC)
following the convention of Girardi and Salaris~(2001).
These corrections,
the inverse effective wavelengths of each passband 
($\micron^{-1}$), and
the adopted reddening law ($A_{\lambda}$; Schlegel et al.~1998)
are all summarized in Table~1.
The results of this work are insensitive
to small uncertainties in the reddening law
or effective wavelengths employed.
Although omitted from Eqn.~1, we also apply a geometric 
correction to our final distance estimate (see below).

We first compare the LMC and
{\it Hipparcos} red clumps 
by assuming $\Delta M_{\lambda}$ = 0.
The resulting apparent
distance moduli are plotted as a function
of inverse effective wavelength in Figure~3
(open circles; error bars omitted),
and listed in Table~1.
A fit for the average
interstellar reddening and
true LMC distance modulus yields 
$E(B-V)$ = $-0.026$ $\pm 0.014_r$
and $(m-M)_0$ = 18.562 $\pm 0.033_r$, which is shown
as the dotted line in Fig.~2.  A negative
reddening correction is unphysical,
and indicates
a significant population difference between the LMC and 
{\it Hipparcos} red clumps.
In order to account for the population difference, we appeal 
to the theoretical calculations of Girardi and Salaris (2001).
These authors generated
artificial LMC and {\it Hipparcos}
red clumps from a grid of core helium-burning
stellar models.  Each clump refers to 
specific star formation and chemical
enrichment histories reported in the literature,
and is thus constrained at least indirectly
by observational data.
Possible systematic errors
associated with the zero-points of
the theoretical models would
cancel in the first order approximation, 
because the population corrections are calculated
only in a relative sense.
This calculation predicts
$\Delta M_I = 0.20$~mag (\S1).
In a private
communication (L.~Girardi \& M.~Salaris~2002),
the same artificial clumps 
yield population corrections of
$\Delta M_V = 0.30$ 
and $\Delta M_K = -0.03$ (Table~1).
The $K$-band absolute magnitude has the smallest correction,
in agreement with the trend noted by Grocholski and Sarajedini (2002).
The resulting modified apparent distance moduli are 
plotted in Fig.~3 (solid circles with error bars),
and listed in Table~1.
Solving for the reddening and true distance modulus yields
$E(B-V)$ = 0.089 $\pm 0.015_r$ and $(m-M)_0$ = 18.506 $\pm 0.033_r$
(solid line in Fig.~3).
The excellent fit of the red clump data 
in Fig.~3 lends indirect support to the 
photometric zero-points adopted in this work.
Applying a small correction for the inclination of
the LMC disk and the location of our fields (\S2), 
the true LMC distance modulus
is $(m-M)_0$ = 18.493 $\pm 0.033_r$.

\section{Conclusion}

We have presented new multi-wavelength luminosity functions
of LMC red clump stars.  Our analysis yields a
distance modulus of $\mu_0 = 18.493 \pm 0.033_r$.
The $K$-band red clump
distance to the LMC is 49.96 $\pm$ 0.77 kpc
in good agreement
with the average published value 
(50.1 kpc; Freedman et al.~2001).
We have also derived an average reddening 
of $E(B-V)$ = 0.089 $\pm 0.015_r$. 
For comparison, the Galactic foreground reddening in this
direction is $E(B-V) = 0.06 \pm 0.02$ (Oestreicher et al.~1995).
Our mean reddening result is consistent
with the foreground reddening, and this is a
success of the Girardi and Salaris (2001) models.

The statistical error associated with our 
new distance result
is due to the
{\it Hipparcos\,} calibration (3\%),
and measuring the LMC red clump brightnesses (1\%).
The overall systematic error may be comparable.
Our photometry is standardized with a 2\% 
systematic uncertainty.
This error is also uncorrelated between $K$ and $V$ or $I$.
However, our zero-point checks do
not support this level of photometric accuracy
unless ground-based comparison data are 
in some cases too bright by up to 0.1 mag.
We have suggested that this may be possible.
Uncertainties in our reddening and geometric corrections
contribute negligible systematic error to the distance result.
The systematic error due to the population correction
is probably also small.  The worst case of making
no correction at all leads to a change in 
modulus of less than 6\%.  However, this also
yields a negative reddening correction.   
For realistic reddenings, the population
correction to the true distance modulus is
approximately the same as the correction in the $K$-band (3\%).
In summary, the systematic errors in our
$K$-band red clump distance to the LMC are probably
of order 2\% (photometric calibration) and
3\% (population correction), and thus comparable to the
statistical error obtained.

The $K$-band red clump distance to the LMC
is in agreement with previously reported $I$-band results (\S1).
The only serious discrepancy is the short distance
result of Udalski (2000) based on OGLE~II data.
Udalski (2000) finds a mean dereddened red clump brightness
of $I_0 = 17.944 \pm 0.014_r$ for 9 fields in the LMC halo.
Those fields are on average $2^{\circ}.12$ from the center of the LMC,
nearly perpendicular to the line-of-nodes, and
on the {\it near} side of the inclined disk (van~der~Marel et al.~2002).
Correcting for the 0.02 mag zero-point offset between OGLE~II
and our {\it HST\,}/WF data, and for geometric projection, 
the Udalski (2000) red clump brightness becomes $I_0 = 18.024$ mag
(at LMC center).
For comparison, our dereddened red clump brightness 
(Table~1) corrected to the
LMC center is $I_0 = 18.019$ mag.  The results of this work,
Udalski (2000), and Romaniello et al.~(2000) therefore all agree,
which lends strong support to the accuracy 
of the $K$-band red clump distance to the LMC.

\begin{acknowledgements}

D.A. acknowledges the referee, K.Z.~Stanek, for his helpful comments,
L.~Girardi and M.~Salaris for their communications,
and A.~Crotts for support (grant No.~AST-00-70882).
D.M.~supported by FONDAP Center for Astrophysics and FONDECYT.
Based on observations made with the NASA/ESA {\it Hubble Space Telescope}, 
obtained at STScI operated by AURA, Inc.~under
NASA contract NAS 5-26555
(GO proposal nos.~5901 \& 7306 to K.C.).
Work performed under auspices of U.S. Dept. of Energy,
National Nuclear Security Administration by UC-LLNL
(contract No.~W-7405-Eng-48).
Use of 2MASS data,
a joint project of the Univ.~of Mass.~and 
IPAC/Caltech funded by NASA and the NSF is acknowledged.

\end{acknowledgements}

\begin{deluxetable}{cccc}
\tablewidth{12cm}
\tablecaption{Red Clump Data}
\tablehead{
\colhead{Quantity} &
\colhead{$K$} &
\colhead{$I$} & 
\colhead{$V$} 
}
\startdata
$\lambda^{-1}$ ($\micron^{-1}$) & 0.48 & 1.24 & 1.81 \\
$A_{\lambda}/E(B-V)$\tablenotemark{A} & 0.35 & 1.96 & 3.24 \\
 & & & \\
$m_{\lambda}$ \tablenotemark{B}  & $16.974\pm0.009$ & $18.206\pm0.009$ & $19.233\pm0.009$ \\
$\sigma_{\lambda}$  & 0.155 & 0.182 & 0.180 \\
$\chi^2$ & 2.2 & 2.1 & 1.6  \\
 & & &  \\
$M_{\lambda}$ \tablenotemark{C}  & $-1.60\pm0.03$ & $-0.26\pm0.03$ & $0.73\pm0.03$ \\
$\sigma_{\lambda}$  & 0.28 & 0.21 & 0.21 \\
$\chi^2$ & 1.4 & 0.9 & 1.7  \\
$(m - M)_{\lambda}$ & 18.57 & 18.47 & 18.50 \\
 & & &  \\
$\Delta M_{\lambda}$\tablenotemark{D} & $-$0.03 & 0.20 & 0.30 \\
$M_{\lambda}$   & $-1.57\pm0.03$ & $-0.46\pm0.03$ & $0.43\pm0.03$ \\
$(m - M)_{\lambda}$ & $18.54\pm0.03$ & $18.67\pm0.03$ & $18.80\pm0.03$ \\
\enddata
\tablenotetext{A}{Adopted reddening law from Schlegel et al.~(1998), which
refers to the effective wavelength of each passband.}
\tablenotetext{B}{Peak brightness and width of the LMC clump,
followed by $\chi^2$ per d.o.f.~of the best-fit model luminosity function.}
\tablenotetext{C}{Peak brightness and width of the {\it Hipparcos\,} clump
(data from Alves 2000),
followed by $\chi^2$ per d.o.f.~of the best-fit model luminosity function.
The LMC apparent distance moduli based on this
calibration are given on the next line (see text).}
\tablenotetext{D}{The population correction for the LMC red clump
absolute magnitude calculated from theoretical models 
(Girardi \& Salaris 2001).  The modified {\it Hipparcos\,}
calibration is given on the next line, 
followed by the resulting LMC 
apparent distance moduli.}
\end{deluxetable}

\clearpage
\begin{figure}
\plotone{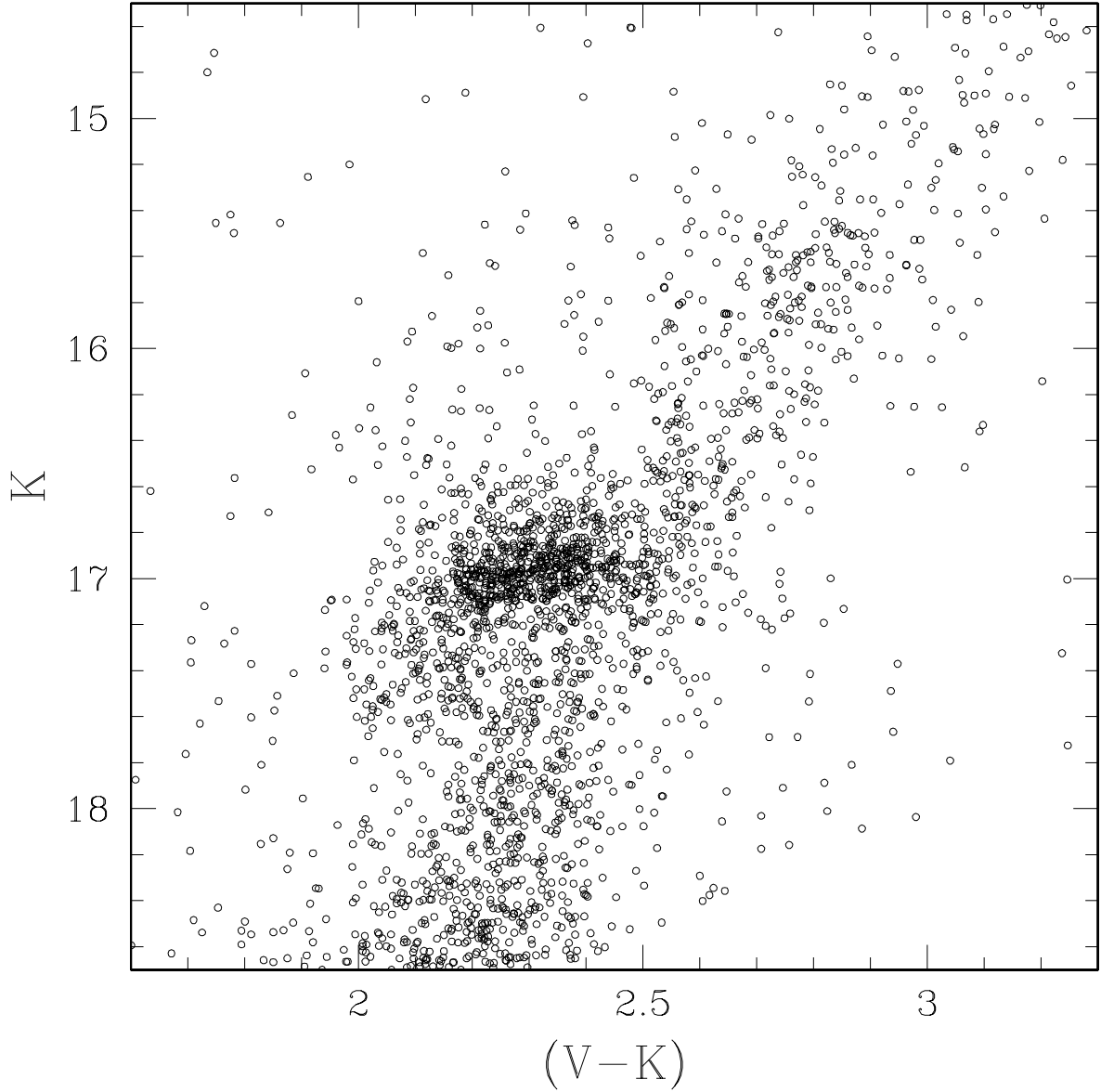}
\caption{The $K$,$(V-K)$ color-magnitude diagram 
of the LMC red clump (2353 stars).
The red clump appears at $K \sim 17$ and lies
mostly blueward of the giant branch.
Error bars on the individual data points are about the
size of the symbols plotted (open circles) at the
brightness of the red clump.}
\end{figure}

\clearpage
\begin{figure}
\plotone{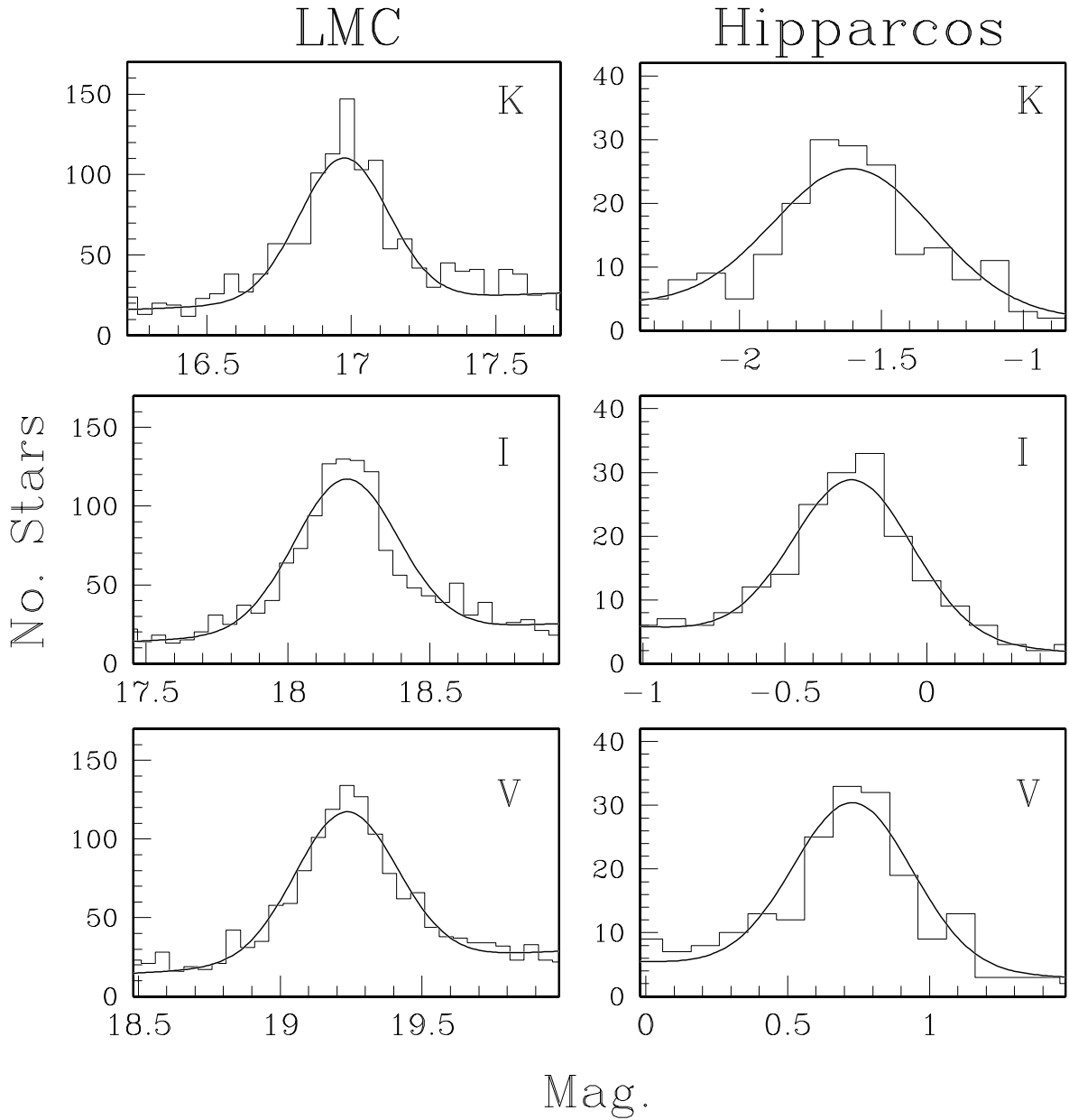}
\caption{The $K$, $I$, and $V$ luminosity functions for 
the LMC red clump are shown in the left panels (bin size 0.05 mag).
In the right panels, the $K$, $I$, and $V$ luminosity functions for
the {\it Hipparcos\,} red clump are shown (bin size 0.1 mag).
The best-fit model functions 
consisting of a Gaussian and a 
linear background are overplotted as solid lines.
These model fits yield
the peak brightness of the red clump in each passband.}
\end{figure}

\clearpage
\begin{figure}
\plotone{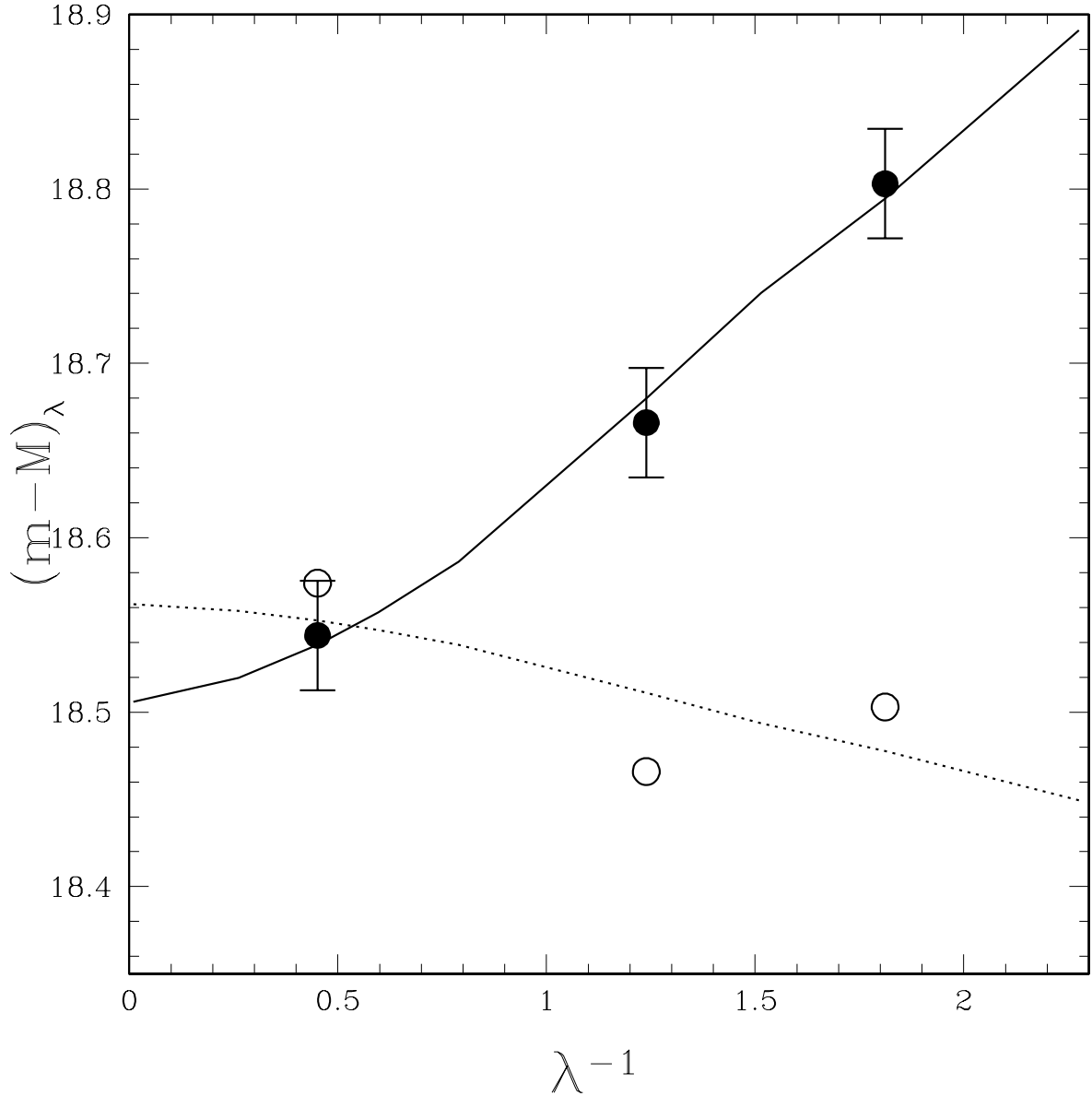}
\caption{Apparent LMC distance moduli plotted as a function
of inverse effective wavelength ($\micron^{-1}$).
The open circles (error bars omitted)
are based on the {\it Hipparcos\,} calibration
without adjustment for population effects.  The
best fit for the mean interstellar reddening correction and
true LMC distance modulus is shown as a dotted line.
The filled circles
(with error bars) are based on the {\it Hipparcos\,} calibration
modified for a mean population difference in the LMC.
The solid line through these data points shows the best-fit 
mean interstellar reddening correction and true LMC
distance modulus:
$E(B-V) = 0.089 \pm 0.015_r$ and $(m-M)_0 = 18.506 \pm 0.033_r$. }
\end{figure}

\end{document}